\documentclass[twocolumn]{article}
\usepackage[top=1in]{geometry}
\usepackage[utf8]{inputenc}
\usepackage{graphicx}
\usepackage{amsmath}
\usepackage{amssymb}
\usepackage{amsfonts}
\usepackage{authblk}
\usepackage{sectsty}
\usepackage{comment}
\usepackage{xcolor}

\sectionfont{\centering}

\title{\begin{flushleft} Entanglement distribution modeling with quantum memories in a global and local clock system \end{flushleft}}

\author[1]{Tasmi R. Ahmed}
\author[1]{Fares Nada}
\author[1]{Amber Hussain}
\author[1]{Connor Kupchak \thanks{\raggedright Corresponding author: connorkupchak@cunet.carleton.ca}\\
TasmiAhmed@cmail.carleton.ca\\
FaresNada@cmail.carleton.ca\\
AmberHussain@cmail.carleton.ca}
\affil[1]{Department of Electronics, Carleton University, Ottawa, ON K1S 5B6, Canada}

\date{}
\begin{document}
\maketitle

\begin{abstract}

We report an innovative model for predicting entanglement distribution between end parties of a quantum network using our in-house simulation algorithm. Our implementation is based on stochastic methods that are built upon a unique global and local clock system for monitoring expectations with finite quantum memory (QM) parameters. This allows us to tabulate rates with independently operating quantum repeater nodes in a distribution chain.  The numerical simulations presented utilize a stochastic modeling of QM efficiency and storage lifetime. The findings presented reveal the translation of the effects of QM lifetime on the spread of time needed for successful entanglement distribution between end parties. Our model based on this transformative clock scheme will make an impactful addition to quantum network simulators platforms.  
\end{abstract}

\section*{{Introduction}}
The heightened interest in establishing the future quantum internet has led to the demonstration of quantum communication testbeds for distributing quantum states between distant parties \cite{Pompili2021, Valivarthi2020, Du2024}. This has encompassed quantum networks based on direct transmission without full-scale repeater hardware over distances of 10$^2$ km~\cite{Ursin2007} in free space and up to 10$^3$ km~\cite{Yin2017} with satellite links as well as memory-compatible nodes in metropolitan area experiments over shorter distance of 10$^1$ km~\cite{Liu2024}.  These advancements have included both qubit-based distribution with a focus on quantum key distribution (QKD) and more powerfully, entanglement distribution for more advanced applications like highly secure entanglement-based QKD schemes~\cite{Eckert91}, sensing~\cite{Shapiro2018}, distributed computing~\cite{Main2024} and long range telescopy~\cite{Gottesman2012}. The development of these rudimentary entanglement-based networks remains challenging due to key hardware limitations, such as source purity and brightness, detector limitations and deterministic loss that occurs in the propagation of quantum channels~\cite{Azuma2023}.  Quantum repeaters and multimode solutions~\cite{Sinclair2014} offer a pathway to overcoming these photonic loss detriments~\cite{Duan2001}. The final, global-scale network will almost certainly contain QM devices~\cite{Sangouard2011} that allow the storage and buffering of qubits such that entanglement can be distributed in an asynchronous fashion~\cite{WangIEEE2022}. 

The future repeater-based network performance will be reliant on many QM factors including memory efficiency, memory related noise and wavelength and bandwidth compatibility. Also vital, the QM coherence time will play a key role in the timing of photons for successful entanglement swapping operation and maintaining high-fidelity entanglement~\cite{Victora2023}. Various platforms are being explored for QM development to effectively operate under this large set of requirements that includes transmission at telecommunication wavelengths, long storage lifetime QMs and a low noise system~\cite{Lei2024}. The development of these networks will occur in incremental steps as they move toward an infrastructure capable of communications on the global scale. 

In light of the technological advancements, there has been a surge of reports on the development of models to predict the performance of future quantum networks~\cite{Bel2025}.  These have taken various forms including modeling of key distribution rates for BB84~\cite{Bennett1984}, full stack quantum network simulators~\cite{DiAdamo2021} and the introduction of stochastic methods to provide a predictive tool. Many network packages are open-source, event-driven simulators designed to model quantum networks, each with a distinct focus and set of capabilities. These range from simulating with errors~\cite{Satoh2022}, full-stack architectures to aid in the design of new network protocols~\cite{Wu2021}, and application specific simulators like quantum key distribution~\cite{Soler2024}

In this work, we present our in-house simulation tool designed to model entanglement distribution in a linear chain of quantum repeaters. Our approach implements a global and local clock system for time simulation. These clocks serve as counters that track the elapsed time steps and memory-ageing against the probabilistic nature of QM storage and lifetimes via Monte Carlo methods~\cite{Satoh2022}. This allows the model to tabulate both instant (synchronous) and delayed (asynchronous) entanglement distribution between parties. Here, we analyze the entanglement generation rates between end-nodes for various QM-based repeater related parameters. Compared to general purpose discrete-event simulators~\cite{DiAdamo2021_2, Satoh2022, Wu2021}, the model presented is a compact, rate-focused layer that estimates memory-limited success statistics through rejection sampling. Further, the clock system can be embedded as a subroutine in cases where full-stack event engines are unnecessary or expensive for parameter scans. Our findings provide network insights as they relate to this clock-based bookkeeping. We believe the model provides a beneficial addition to the existing set of network simulators~\cite{Bel2025}, where such software will become increasingly important as we move closer to a quantum internet~\cite{Wehner2018}.

\section*{{Entanglement distribution schemes}}
\begin{figure*}[t]
  \centering
  \includegraphics[width=1\linewidth]{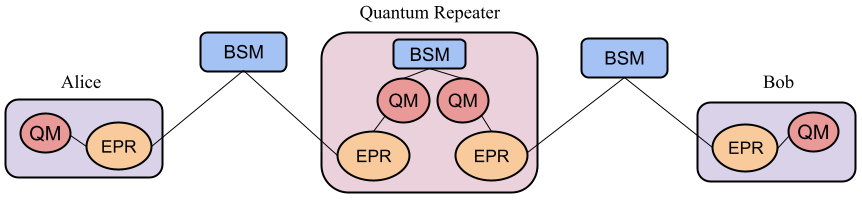}
  \caption{Entanglement distribution chain with a single midpoint quantum repeater (QR)  and intermediate Bell-state measurement (BSM) stations. Alice and Bob each hold an Einstein--Podolsky--Rosen (EPR) entangled-photon source and a quantum memory (QM). The central node contains two EPR sources, two QMs and a central BSM. Intermediate BSMs lie halfway between each end node and the centre of the distribution chain. With a single midpoint quantum repeater, photons propagate along fibre segments of length $L/4$ between an end node and an intermediate BSM, and between an intermediate BSM and the centre yielding total distance $L$. Lines indicate the optical paths.}
  \label{fig:figure1}
\end{figure*}

Entanglement distribution protocols are a set of procedures, which together, act as a delivery service to aid in the establishment of quantum entanglement between two distant locations. However, doing so over long distances is exceptionally challenging due to deterministic photonic loss that plagues long-distance channels. The recognized solution to circumvent this problem is the introduction of photonic quantum repeater (QR)~\cite{Briegel1998} to reduce photon transmission distance. 

A schematic diagram of this swapping configuration is shown in Fig.~\ref{fig:figure1}. Here, Alice and Bob each have an Einstein-Podolsky-Rosen (EPR)~\cite{Einstein1935} source and a QM. The central node,
which acts as a single quantum repeater, contains two EPR sources, two QMs and a bell state measurement (BSM) apparatus. Additionally, there exist two intermediate BSM devices situated midway between Alice/Bob and the central repeater node.

The system's operation begins with all four EPR sources generating entangled photon pairs. Alice (and Bob) send one photon from their respective pairs toward the intermediate BSM, while storing the other photon in their QM.  Simultaneously, the two EPR sources at the central node send one photon from each of their pairs to the intermediate BSMs, holding the other photons in their dedicated QMs.

Upon the arrival of photons at the intermediate BSMs, an entanglement swapping operation is performed. The exact composition of the BSM depends on the degrees of freedom for the entangled state. In our model we will assume polarization entanglement so the BSMs will employ a combination of single-photon detectors and beam splitters to register the detection patterns that signify a successful swapping operation. The swapping operation erases the "which-path" information and establishes entanglement between the QMs held by Alice and Bob and the QMs at the central repeater node. Once the central node's QMs confirm that they each hold an entangled photon, they release them to the central BSM apparatus. A successful BSM here projects Alice and Bob's QMs into an entangled state with each other. A notification is then sent to Alice and Bob, confirming the successful creation of an end-to-end entangled link. If any step in this sequence fails, the entire process must be repeated. This repeater chain can be extended to a larger number sufficient to overcome the exponential loss in direct transmission.

With the availability of QRs, there are two primary approaches to entanglement distribution: synchronous and asynchronous. In the synchronous case, also known as the Barrett-Kok protocol~\cite{Kok2005}, entanglement must be successfully generated across every link in the repeater chain simultaneously. If a single link fails, the entire process is discarded, and all links must restart. While conceptually simple, this approach is highly inefficient over long distances because the probability of all links succeeding simultaneously decreases polynomially with the number of repeater nodes.

In the asynchronous scenario~\cite{WangIEEE2022}, each repeater link attempts to generate entanglement independently. Once a link succeeds, the entangled states are stored in the QMs associated with the link, until the remaining links have also succeeded.  This approach dramatically increases the overall probability of success as it avoids discarding links that have been successful. 

The asynchronous aspect provides a loss advantage but places a strict demand on the QM related factors. This includes the memory coherence time~\cite{Shaham2022} which determines how long a quantum state can be held in the QM before succumbing to significant decoherence. This is quantified by the parameter $\tau_{\scriptscriptstyle mem}$ and follows an exponential relationship as $\sim e^{-t/ \tau_{\scriptscriptstyle mem}}$.  Other parameters that were taken into consideration are inefficiencies in the detectors, bell state measurement devices and the memory efficiency. In this work, the group velocity of $v = 2\times10^5$~km/s is assumed which is common for standard silica fibres used in telecommunications where $v \approx c/n_{\mathrm{g}}$ and $n_{\mathrm{g}}\!\sim\!1.5$. A fibre attenuation coefficient $\alpha$ corresponding to 0.2~dB/km, is also assumed which is typical for state-of-the-art fibre at telecommunication wavelengths. Physically, attenuation corresponds to the probability for a single photon to survive a fibre segment of length $\ell$ without being lost. This attenuation scales as $e^{-\alpha \ell}$ and length of fibre will introduce multiplicative transmission losses.

\section*{{Single-link entanglement distribution}}

We begin our analysis for the case of single-link entanglement distribution. Here, both Alice and Bob have a local QM and EPR source and are separated by total distance $L$ with a central BSM apparatus situated at the midpoint. Alice and Bob will synchronously send one photon from their entangled EPR pair to the central BSM station at a correlated arrival time. Alice and Bob's QMs will store their entangled photons for the time necessary for the pair photon to travel to the central BSM and the notification signal to be returned to Alice or Bob corresponding for a total time of $t=2\times(L/2v)$.  In addition to the QM coherence time, the distribution rate is also impacted by the fibre transmission loss experienced by each of Alice and Bob's photons $\sim e^{- \alpha\frac{L}{2}}$, QM storage efficiency $\eta_{\scriptscriptstyle QM}$ and the BSM measurement efficiency $\eta_{\scriptscriptstyle BSM}$ which is typically set as 50\% for the case of a linear optical elements.

Eqns.~(\ref{eq:singlelinkdet})--(\ref{eq:multilinkstoch}) show that end-to-end rate scales as attempt frequency multiplied by per-link success~\cite{Briegel1998, Sangouard2011}. For a single link, the success probability can be shown proportional as $P_{\mathrm{succ}} \propto \eta_{\scriptscriptstyle BSM}\eta_{\scriptscriptstyle D}^2 e^{-\alpha L/2} e^{-\alpha L/2}\, \eta_{\scriptscriptstyle QM}^2\, e^{-t/\tau^A_{\mathrm{mem}}} e^{-t/\tau^B_{\mathrm{mem}}}$; inclusive of detector efficiencies $\eta_{\scriptscriptstyle D}$ and memory storage and coherence factors. In a single photonic entanglement attempt over distance $L$, the success rate scales as $R \sim (v/L)\,P_{\mathrm{succ}}$ up to protocol-dependent prefactors as shown by Eq.~(\ref{eq:singlelinkdet}). The progression to Eqn.~(\ref{eq:singlelinkstoch}) then introduces stochastic events by replacing the deterministic exponential QM factors with Monte Carlo estimates of the same Bernoulli (binary) events in the limit of large numbers ($N_{\mathrm{trials}}\to\infty$). Finally, Eqns.~(\ref{eq:multilinkdet}) and~(\ref{eq:multilinkstoch}) extend the same product of terms to $n$ elementary links of length $L/(2n)$. For the synchronous protocol, we evaluate these expressions numerically.

The deterministic expression for the entanglement distribution rate is given by 
\begin{equation} \label{eq:singlelinkdet}
\begin{split}
  &R^1_{det}(L) = \\ &\frac{v}{L}\eta_{\scriptscriptstyle BSM}\left(\eta_{\scriptscriptstyle QM}\eta_{\scriptscriptstyle D}e^{- \alpha \frac{L}{2}}\right)^2\left(e^{-t / \tau^A_{\scriptscriptstyle mem}}\right)\left(e^{-t / \tau^B_{\scriptscriptstyle mem}}\right).
  \end{split}
\end{equation}
In Eq.~\ref{eq:singlelinkdet} we assume that the QM efficiency $\eta_{\scriptscriptstyle QM}$ and the loss in the channel to the central BSM for each of Alice and Bob channels are equal.

We can now alter Eq.~\ref{eq:singlelinkdet} to include stochastic processes pertaining to the QMs. Now the probability the photon is stored in the QM and the probability each photon is still held after notification from the central BSM is sampled via Monte Carlo methods.  More specifically, in each realization, our model generates two pairs of random variables between ${0,1}$. One of these pairs is allocated to the QM efficiencies of each of Alice and Bob.  Individually, the random numbers are checked to lie below the $\eta_{\scriptscriptstyle QM}$ of Alice or Bob's QM and counted as a success if true. The two additional random variables are individually compared to lie below the lifetime decay curve of Alice or Bob's QM. This process is then repeated over a large number of trials, with successes being individually recorded for each QM efficiency and retainment of optical storage. The number of recorded successes was then divided by the total number of trials and the product of the four statistical means provided the stochastic entanglement distribution rate at a given length $L$. The stochastic distribution rate is given by 

\begin{equation}
  \label{eq:singlelinkstoch}
  \begin{split}
  &R^1_{stoch}(L) = \frac{v}{L}\eta_{\scriptscriptstyle BSM}\eta_{\scriptscriptstyle D}^2\left(e^{- \alpha \frac{L}{2}}\right)^2 \times
   \\ & \left\langle\frac{N^A_{\scriptscriptstyle mem}}{N_{\scriptscriptstyle trials}}\right\rangle
  \left\langle\frac{N^B_{\scriptscriptstyle mem}}{N_{\scriptscriptstyle trials}}\right\rangle
  \left\langle\frac{N^A_{\scriptscriptstyle coh}}{N_{\scriptscriptstyle trials}}\right\rangle
  \left\langle\frac{N^B_{\scriptscriptstyle coh}}{N_{\scriptscriptstyle trials}}\right\rangle.
  \end{split}
\end{equation}

For the stochastic version, we now replace the deterministic dependencies with mean average values $\langle \cdot\rangle$ calculated by the total number of successes divided by the total number of trials $N_{\scriptscriptstyle trials}$. This was implemented for both the successful number of QM storage events $N^{A/B}_{\scriptscriptstyle mem}$ and number of events where the QMs successfully held onto the photons $N^{A/B}_{\scriptscriptstyle coh}$. The binary nature of our parameters with Bernoulli success probability $p$, the sample mean has standard error $\sqrt{p(1-p)/N_{\scriptscriptstyle trials}}$. For the case of $N_{\scriptscriptstyle trials}=10^6$, the relative uncertainty at $p\sim 0.5$ is below $0.2\%$, while at $p\sim 10^{-6}$ the absolute error remains $\sim 10^{-3}$ in $p$, which can appear as visible ripples on logarithmic plots when the rate itself is extremely small.

\begin{figure}[t]
  \centering
  \includegraphics[width=1\linewidth]{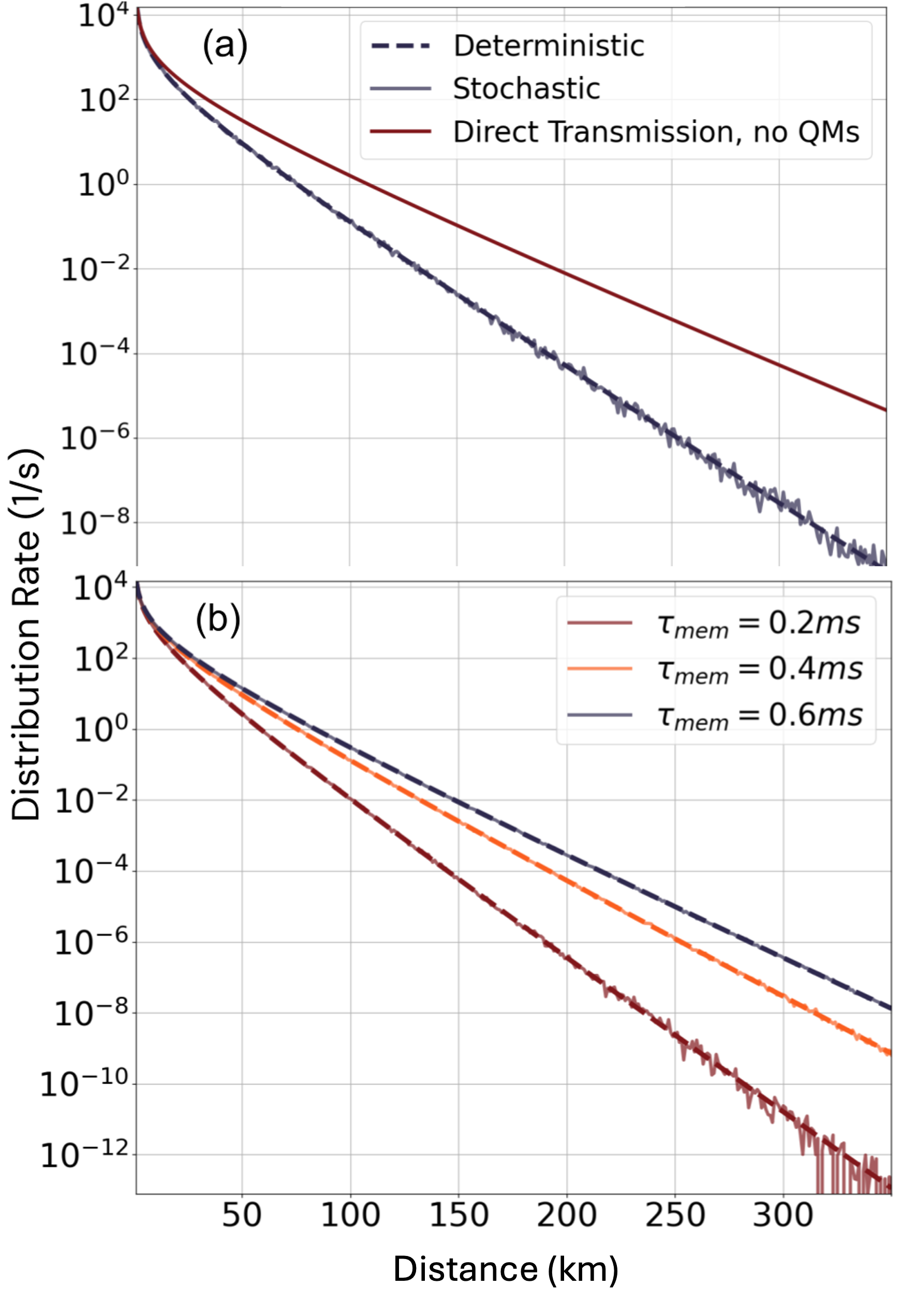}
  \caption{(a) Single link analysis with Alice and Bob each having a local QM and awaiting notification of a successful BSM. Horizontal axis: end-to-end fibre distance $L$ (km). Vertical axis: entanglement distribution rate $R$ (s$^{-1}$). The plot compares the deterministic model (dashed) to the Monte Carlo estimate (solid). The red curve is direct transmission without end memories. (b) As in (a) with horizontal axis $L$ (km) and vertical axis $R$ (s$^{-1}$), for memory coherence times $\tau_{\scriptscriptstyle mem}=0.2, 0.4, 0.6$~ms. Ripples at the lowest rates are finite-sample Monte Carlo fluctuations (see text).}
  \label{fig:singlelink}
  
\end{figure}

Figure~\ref{fig:singlelink} shows the agreement between the single link deterministic (Eq.~\ref{eq:singlelinkdet}) and stochastic (Eq.~\ref{eq:singlelinkstoch}) entanglement rate distributions for a set of QM coherence times. This figure demonstrates how even slight changes in coherence times can drastically alter the distribution rate due to the exponential nature of the lifetime decay. We can also see the increased presence of statistical fluctuations with low probability events corresponding to the shortest QM coherence times.  Also plotted is the case where Alice and Bob transmit directly to the central BSM but do not hold a QM device. In this scenario, the direct transmission yields higher distribution rates since there is an absence loss due to QM efficiency. The single link case serves as a step towards simulation of a multi-link network described in the next sections.

The QM coherence times $\tau_{\scriptscriptstyle mem}$ used throughout this study (e.g.\ 0.1--1~ms) were chosen to span the order-of-magnitude scales reported in QM experiments and reviews~\cite{Shaham2022, WangQu2022} by association highlight scaling laws in QM repeater chains topologies.

On a log-linear scale, Fig.~\ref{fig:singlelink} resembles curves characteristic of coherent optical communications, where the bit-error ratio falls exponentially with received power. In this case, the errors would be dominated by loss and memory failure opposed to classical limitations. As distance increases, multiplicative success probabilities drive the rate downward as success must occur on every required channel and memory element. At the lowest displayed rates, the transmission rates are dominated by Monte Carlo noise and in this treatment are presented to illustrate the asymptotic scaling of our model.

\section*{{Stochastic modeling based on a global and local clock system}}

With the introduction of quantum memories comes the opportunity of both synchronous and asynchronous entanglement distribution methods; timing constraints in protocols are often discussed in terms of clocks~\cite{Shi2022}. Synchronous distribution requires all links to be established simultaneously. Asynchronous allows for storage of entangled photons in their respective QMs, allowing for all the other non-established links to be successful before end-to-end entanglement is achieved leading to higher distribution rates and longer distances.

To achieve stochastic modeling of both synchronous and asynchronous distribution rates, our model implements a global and a local clock system. The global clock tracks the total simulated time per trial, while each local clock tracks how long a given memory has held its excitation for the purpose of sampling decoherence. The clocks serve as a counting management system to track asynchronous photon capture with the stochastic nature stemming from the QM efficiency and coherence time while the clocks themselves are assumed noiseless.

\begin{figure*}[t!]
  \centering
  \includegraphics[width=.9\linewidth]{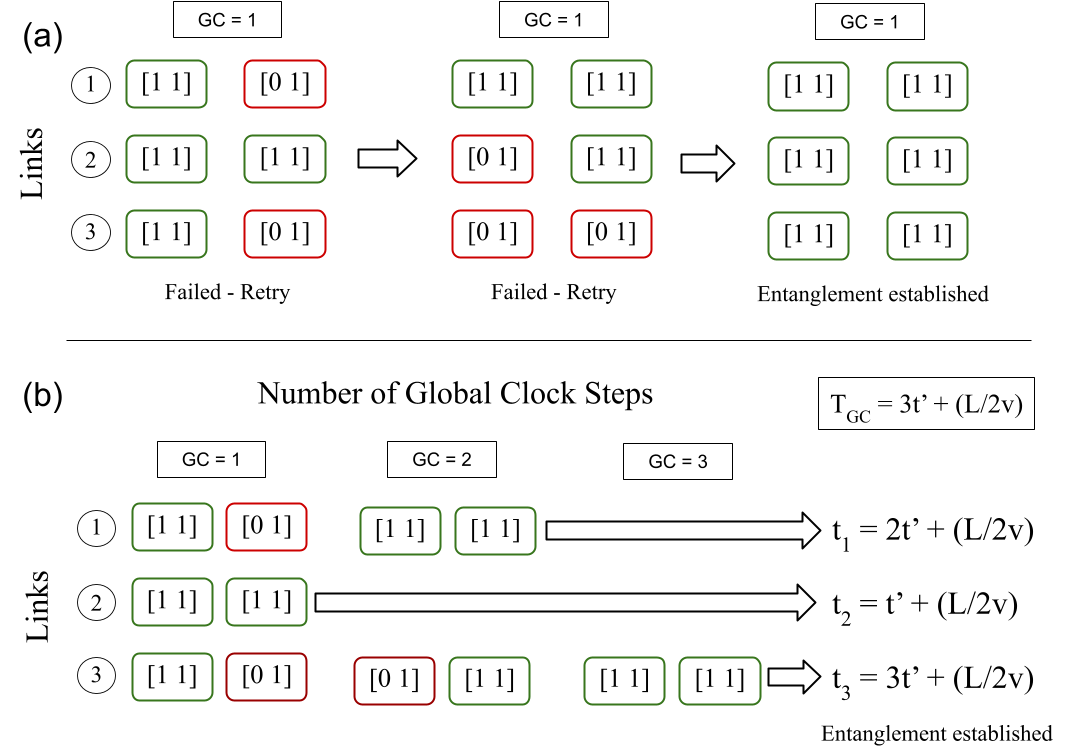}
  \caption{Schematic diagram of the global and local clock scheme for distributing entanglement between Alice and Bob. Each QM keeps a pair of values $[Succ, t_{\mathrm{LC}}]$ corresponding to a binary success indicator and local clock time in units of $t'=L/(2vn)$ for $n$ links. The global clock $T_{\mathrm{GC}}$ advances the total simulation while the local clocks age stored excitations until decoherence or completion. (a) In synchronous attempts all links must succeed in the same global step. (b) In asynchronous attempts the links proceed independent of one another. The variable $T_{\mathrm{GC}}$ records elapsed steps until all links have succeeded.}
  \label{fig:clockscheme}
\end{figure*}

A schematic of our scheme used to simulate synchronous entanglement distribution is shown in Fig.~\ref{fig:clockscheme}(a). Here, we show a three-link system, where each link contains two quantum memories. Each QM is represented by a rectangle containing a two-element array. The first element holds a binary value corresponding to a successful ('1') or failed ('0') storage attempt, while the second element tracks the local clock time.

To account for the communication time between a node and source, each QM's local clock time will begin at one unit upon storage. Here, a single time unit corresponds to $t'=L/(2vn)$, where $n$ is the number of links. If a QM records a successful storage event, its element values will be set to [1 1], corresponding to a binary value for storage and a local clock time of 1. Since we are restricted to the case of synchronous entanglement distribution, the global clock time will always have a value of one time unit when entanglement is established.

For asynchronous entanglement distribution, the schematic is shown in Figure~\ref{fig:clockscheme}(b) where we must now track the net global clock time in addition to the local clock for each QM. Similar to the synchronous scheme, each QM contains a two-element array corresponding to a successful storage event and the local clock time. For each QM-link pair, the local clocks begin counting once a storage event has been stored in each QM of the pair. There is always a chance that the stored entangled state may be lost due to decoherence, so the local clock of each QM is individually sampled against a memory coherence clock at every global clock step. If the entanglement is lost due to decoherence, that link fails and must be re-established. The protocol continues and the global clock continues to run until all links possess entanglement. Once established, the recorded global clock time $T_{GC}$ corresponds to the longest local clock time still active plus the communication time from the midpoint of the chain to an end-party $L/(2v)$. The only time an asynchronous establishment attempt fails is when entanglement fails across all links simultaneously; otherwise, the global clock will continue to run until all links have successfully achieved entanglement.

\section*{Entanglement distribution with synchronous links}

In a quantum repeater network, synchronous entanglement generation, also known as the Barrett-Kok protocol, is a method where all repeater nodes attempt to swap entanglement simultaneously. The process is only considered a success if every link is established at the same time. If one link fails, the entire attempt is discarded and the whole sequence must be repeated. While this approach ensures the entanglement is the most current, it is inefficient over long distances because the probability of all links succeeding at once is low. However, this scheme allows intermediate nodes to act as quantum routers, enabling a quantum network without needing direct fiber connections between every node pair. A synchronous protocol is best suited for hardware with short quantum memory lifetimes, as it avoids the long wait times required by other methods.

To calculate the entanglement distribution rates we extend  Eqs.~\ref{eq:singlelinkdet} and~\ref{eq:singlelinkstoch} to account for multiple links $n$.  For the deterministic rate this yields

\begin{equation} \label{eq:multilinkdet}
\begin{split}
  &R^n_{det}(L) = \frac{2vn}{L}\eta^2_{\scriptscriptstyle QM}\left(\eta_{\scriptscriptstyle BSM}\left(\eta_{\scriptscriptstyle D}e^{- \alpha \frac{L}{2n}}\right)^2\right)^n \times \\ &\left[\eta_{det} e^{-t /(\tau^A_{\scriptscriptstyle mem}+\tau^B_{\scriptscriptstyle mem})}\right]^{n-1}, 
  \end{split}
\end{equation}

\noindent with $\eta_{det}=\eta_{\scriptscriptstyle BSM}\eta_{\scriptscriptstyle QR}\eta^2_{\scriptscriptstyle QM}\eta^2_{\scriptscriptstyle D}$ corresponding to the total loss experienced inside a QR node. The factor $\eta_{\scriptscriptstyle QR}$ considers the total efficiency of the QR (coupling efficiencies etc.) and is taken to be a value of 0.5 in all cases. Here, we can see that the synchronous requirement will negate the benefit due to shorter transmission links. For simplicity of our model we assume a fixed delay between each attempt that is dictated by the travel time from the sources and BSM apparatus between node stations. Note that we will not use Eq.~\ref{eq:multilinkdet} explicitly in our analysis but it will serve as a reference equation for extending the stochastic case of Eq.~\ref{eq:singlelinkstoch} for multiple links to

\begin{equation}
  \label{eq:multilinkstoch}
  \begin{split}
  &R^n_{stoch}(L) = \frac{2vn}{L}\eta^2_{\scriptscriptstyle QM}\left(\eta_{\scriptscriptstyle BSM}\left(\eta_{\scriptscriptstyle D}e^{- \alpha \frac{L}{2n}}\right)^2\right)^n \times
   \\ &\left[\eta_{\scriptscriptstyle stoch}\left\langle\frac{N^A_{\scriptscriptstyle mem}}{N_{\scriptscriptstyle trials}}\right\rangle
  \left\langle\frac{N^B_{\scriptscriptstyle mem}}{N_{\scriptscriptstyle trials}}\right\rangle
  \left\langle\frac{N^A_{\scriptscriptstyle coh}}{N_{\scriptscriptstyle trials}}\right\rangle
  \left\langle\frac{N^B_{\scriptscriptstyle coh}}{N_{\scriptscriptstyle trials}}\right\rangle \right]^{n-1},
  \end{split}
\end{equation}
\noindent where $\eta_{stoch}=\eta_{\scriptscriptstyle BSM}\eta_{\scriptscriptstyle QR}\eta^2_{\scriptscriptstyle D}$. 
\begin{figure*}[t]
  \centering
  \includegraphics[width=1\linewidth]{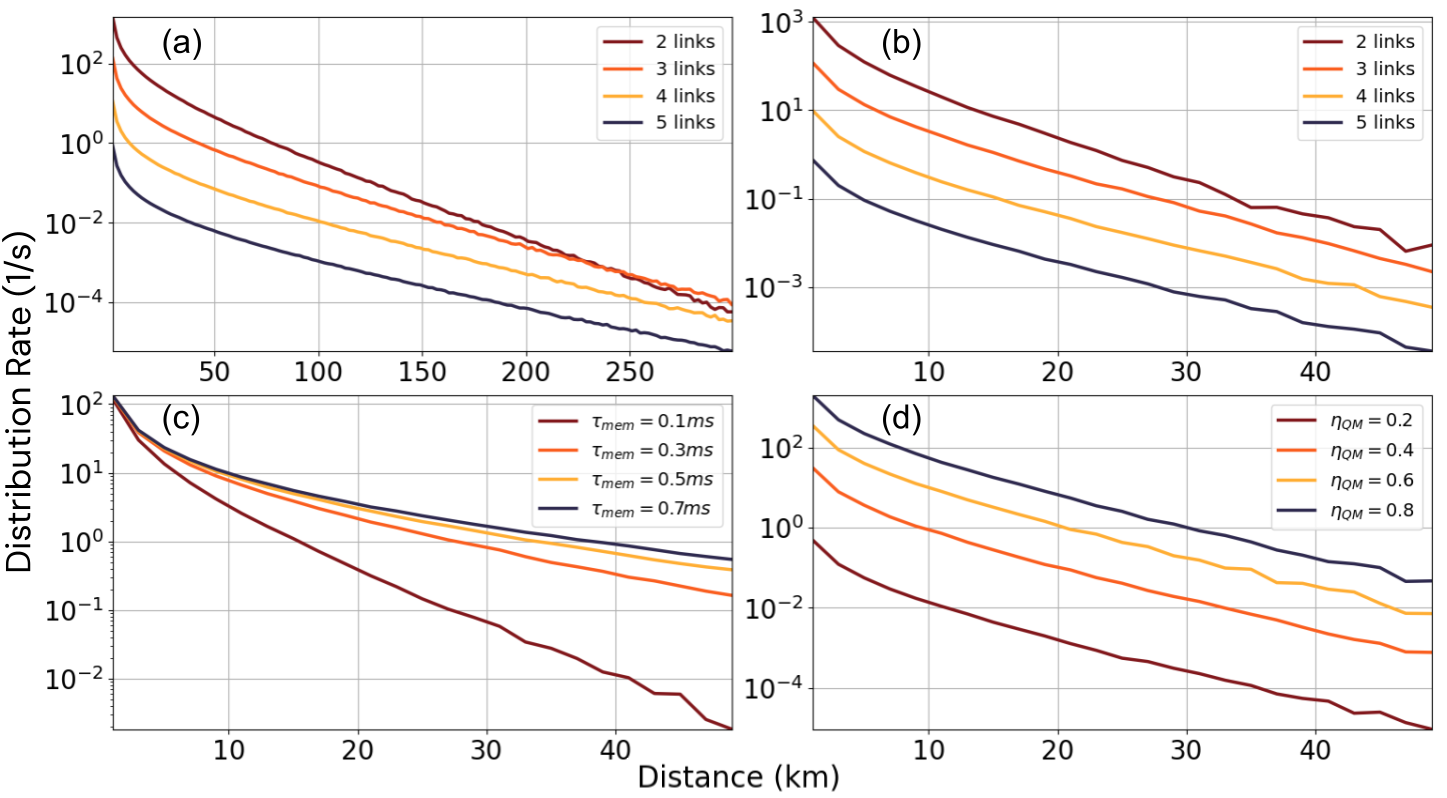}
  \caption{Entanglement distribution rates versus distance for synchronous link establishment: 
  (a) $\tau_{\scriptscriptstyle mem} = 1$ ms, $\eta_{\scriptscriptstyle QM}$ = 0.5, varying number of links;
  (b) $\tau_{\scriptscriptstyle mem} = 0.1$ ms, $\eta_{\scriptscriptstyle QM}$ = 0.5, varying number of links;
  (c) varying $\tau_{\scriptscriptstyle mem}$, $\eta_{\scriptscriptstyle QM}$ = 0.5, number of links = 3;
  (d)  $\tau_{\scriptscriptstyle mem} = 0.1$ ms, varying $\eta_{\scriptscriptstyle QM}$, number of links = 3.}
  \label{fig:synchdist}
\end{figure*}

Figure~\ref{fig:synchdist} illustrates Eq.~\ref{eq:multilinkstoch} for varying efficiencies, number of links and coherence times, and shows the impact for different factors highlighting the contrast between exponential and polynomial losses. As in Fig.~\ref{fig:singlelink}, the log--linear slopes echo the intuition from coherent optical links: incremental loss in each segment multiplies the end-to-end success probability, analogous to how error probabilities compound in classical transmission, though here the dominant ``errors'' are photon loss and failed BSMs. Fig.~\ref{fig:synchdist}(a) shows the rate dependencies assuming an infinite QM coherence time and $\eta_{\scriptscriptstyle QM}=$0.5 for a varying number of links. It can be seen that at low distances, polynomial losses yield the greatest impact. At distances near 200~km the impact of reduced exponential losses from links takes over causing the rate lines to cross over. We extend this analysis for a fixed QM coherence time in Fig.~\ref{fig:synchdist}(b) for $\tau_{\scriptscriptstyle mem}=0.1$ ms, which yields similar results to the infinite coherence time case as little photon loss in the memories occurs at these distances. Figs.~\ref{fig:synchdist}(c) and~\ref{fig:synchdist}(d) display the rates with varying coherence times and memory efficiencies for a three-link ($n=3$) chain with one central repeater node.  

In synchronous entanglement generation, non-unity coherence times and memory efficiencies cause the final rate to fall below that of direct transmission. For this reason, a lower number of nodes and links increases the probability of success. Because the overall link success depends on the synchronization of all repeater nodes, the time for a successful entanglement distribution always corresponds to a single global clock unit.

This demonstration of the synchronous protocol not only proves our algorithm's ability to simulate a single chain repeater architecture, but more importantly, establishes the local/global clock framework for extending our algorithm to more complex repeater topologies.

\section*{Entanglement Distribution Using Asynchronous Links}

The benefit of our global/local clock can be highlighted in the case where the links of the QR chain are capable of asynchronous operation. Alice, Bob or the QR stations send a photon from an entangled EPR pair to the central BSM and the other photon from the EPR is then held in a dedicated QM. Our simulation accounts for link-by-link entanglement times by using a global and local clock system to test entanglement against two probabilistic factors: first, the probability that entanglement was established at a given link and second, the probability that the entanglement was sustained while other links establish entanglement. The latter is modeled using a lifetime decay exponential tied to the local clock time of each established link. A trial continues as long as one link still holds entanglement and until every link has successfully established entanglement. For asynchronous generation, the QMs in each link are tested at each global clock unit; thus, if all links fail at the same instant, the trial is recorded as a failure.

More elaborate repeater strategies can include sequential pumping, parallel heralding across multiple spatial or spectral modes, or nested entanglement swapping~\cite{Briegel1998}. These strategies would change scheduling and resource allocation but still, at its core, reduce sampling link successes and the corresponding QM lifetimes. The present implementation focuses on a single linear topology with either a synchronous limitation or asynchronous completion. Extending our treatment to nested layers would entail maintaining a local clock per segment at each nesting depth, and strategies would scale the attempt rate with the number of independent connections and will be expanded upon in future studies.

\begin{figure*}[t]
  \centering
  \includegraphics[width=1\linewidth]{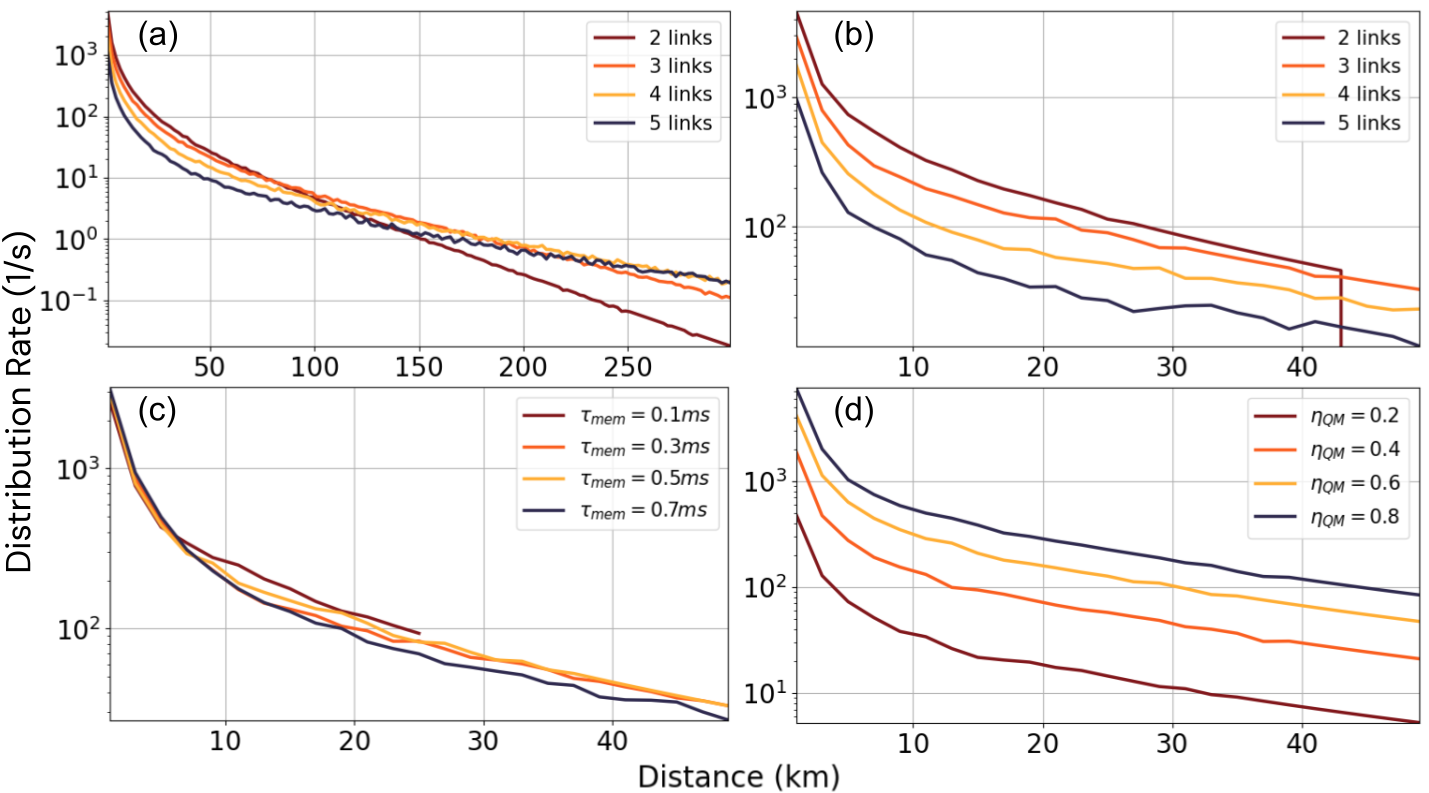}
  \caption{Entanglement distribution rates versus distance for asynchronous link establishment where the rate is divided by $\langle T_{\scriptscriptstyle GC}\rangle$, the mean global clock time for when all links are successful:
  (a) $\tau_{\scriptscriptstyle mem} = 1$ ms, $\eta_{\scriptscriptstyle QM}$ = 0.5 for different number of links $n$;
  (b) $\tau_{\scriptscriptstyle mem} = 0.1$ ms, $\eta_{\scriptscriptstyle QM}$ = 0.5 for different number of links $n$;
  (c) varying $\tau_{\scriptscriptstyle mem}$, $\eta_{\scriptscriptstyle QM}$ = 0.5 and $n=3$;
  (d)  $\tau_{\scriptscriptstyle mem} = 0.1$ ms, varying $\eta_{\scriptscriptstyle QM}$, and $n=3$.}
  \label{fig:asynchrondist}
  
\end{figure*}

Here, the successful BSM events do not occur in a dedicated time interval and results are tabulated at any time corresponding to successful entanglement distribution. The total distribution rate for the scheme $R_{\scriptscriptstyle Asynch}$ is given by

\begin{equation}\label{eq:asynchrate}
  \begin{split}
  &R^n_{\scriptscriptstyle Asynch}(L) = \frac{2vn}{ \langle T_{\scriptscriptstyle GC}\rangle L}\eta^2_{\scriptscriptstyle QM}\eta_{\scriptscriptstyle BSM}\left(\eta_{\scriptscriptstyle D}e^{- \alpha \frac{L}{2n}}\right)^2 \times
   \\ &\left[\eta_{\scriptscriptstyle stoch}\left\langle\frac{N^A_{\scriptscriptstyle mem}}{N_{\scriptscriptstyle trials}}\right\rangle
  \left\langle\frac{N^B_{\scriptscriptstyle mem}}{N_{\scriptscriptstyle trials}}\right\rangle
  \left\langle\frac{N^A_{\scriptscriptstyle coh}}{N_{\scriptscriptstyle trials}}\right\rangle
  \left\langle\frac{N^B_{\scriptscriptstyle coh}}{N_{\scriptscriptstyle trials}}\right\rangle \right]^{n-1}.
  \end{split}
\end{equation}
Eq. \ref{eq:asynchrate} is derived by removing synchronous transmission requirement over all $n$ links and modifying the conventional rate to account for the mean global clock time over all successes, $\langle T_{\scriptscriptstyle GC}\rangle$.

This asynchronous approach can permit a greater probability of a successful end-to-end link compared to synchronous protocols and tabulation of complex stochastic cases. However, this places a more stringent requirement on the coherence time of the QM, but allows circumvention of extended fibre losses. As a result, the effective attempt frequency is reduced, but the overall entanglement generation rate is significantly more than the synchronous scenario and ultimately enables longer communication distances.

The total entanglement rates yielded by Eq.~\ref{eq:asynchrate} are shown in Fig.~\ref{fig:asynchrondist}.  Overall, asynchronous rate exceeds that of the synchronous case by a factor for about $\sim 10^2$ at a shorter range distance near 10 km and a rate improvement of $\sim 10^3$ at distances $>$30 km attributed to the reduction in loss experienced by propagation through fibre.

The utility of our stochastic approach serves as a powerful complement to existing quantum network simulators. This model can be extended beyond single chain networks with entanglement being generated asynchronously in various link configurations. These tools are key for the modeling of future entanglement-based quantum network infrastructures and designing and optimizing quantum hardware.

\section*{Analysis into global and local clock behavior}

\begin{figure*}[t!]
  \centering
  \includegraphics[width=1\linewidth]{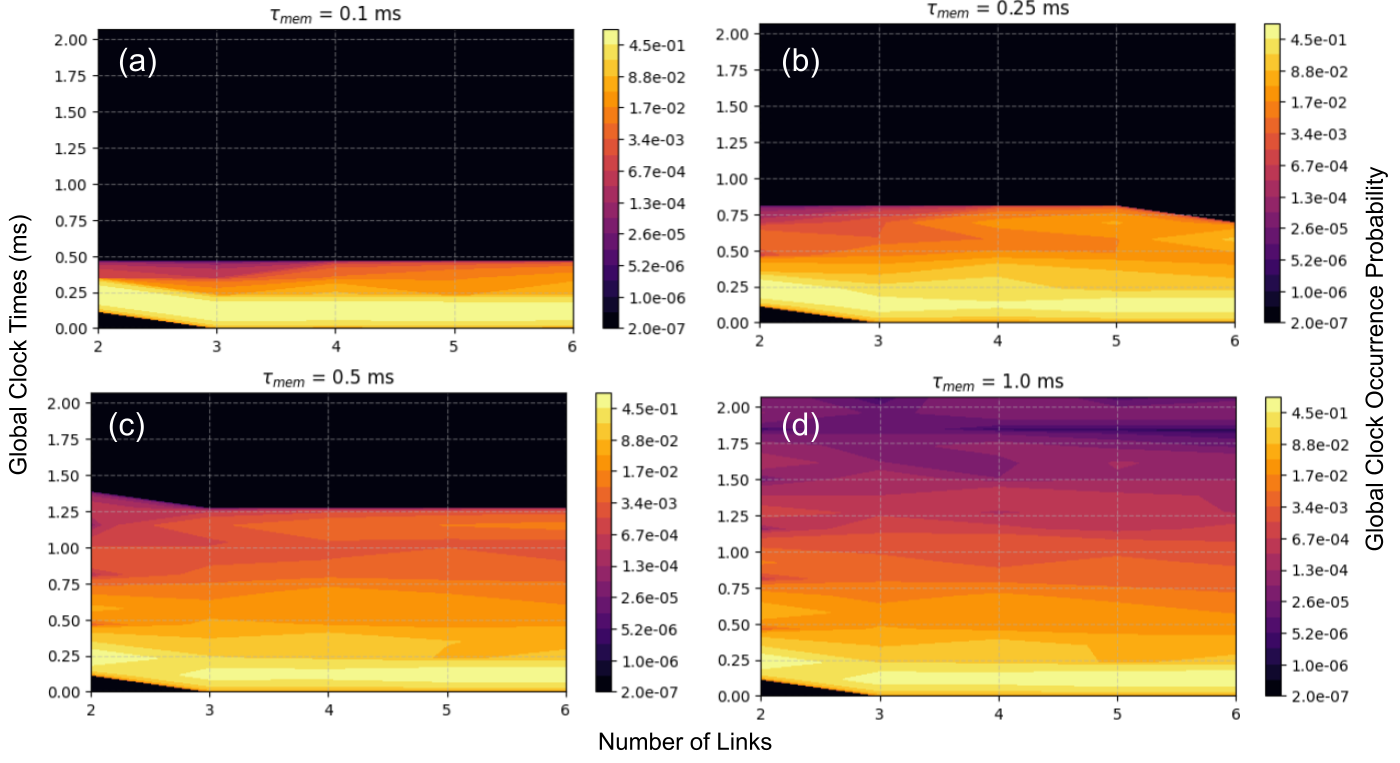}
  \caption{Histograms of global clock time $T_{\mathrm{GC}}$ (simulation steps until all links succeed, conditioned on success) for 1000 successful asynchronous trials. Columns (a)-(d): $\tau_{\scriptscriptstyle mem}=0.1$, 0.25, 0.5, 1.0~ms. Horizontal axis: number of links; vertical axis: $T_{\mathrm{GC}}$; Contours are the success probability (log scale).}
  \label{fig:GCtimevslinks}
  
\end{figure*}

\begin{figure*}[t!]
  \centering
  \includegraphics[width=1\linewidth]{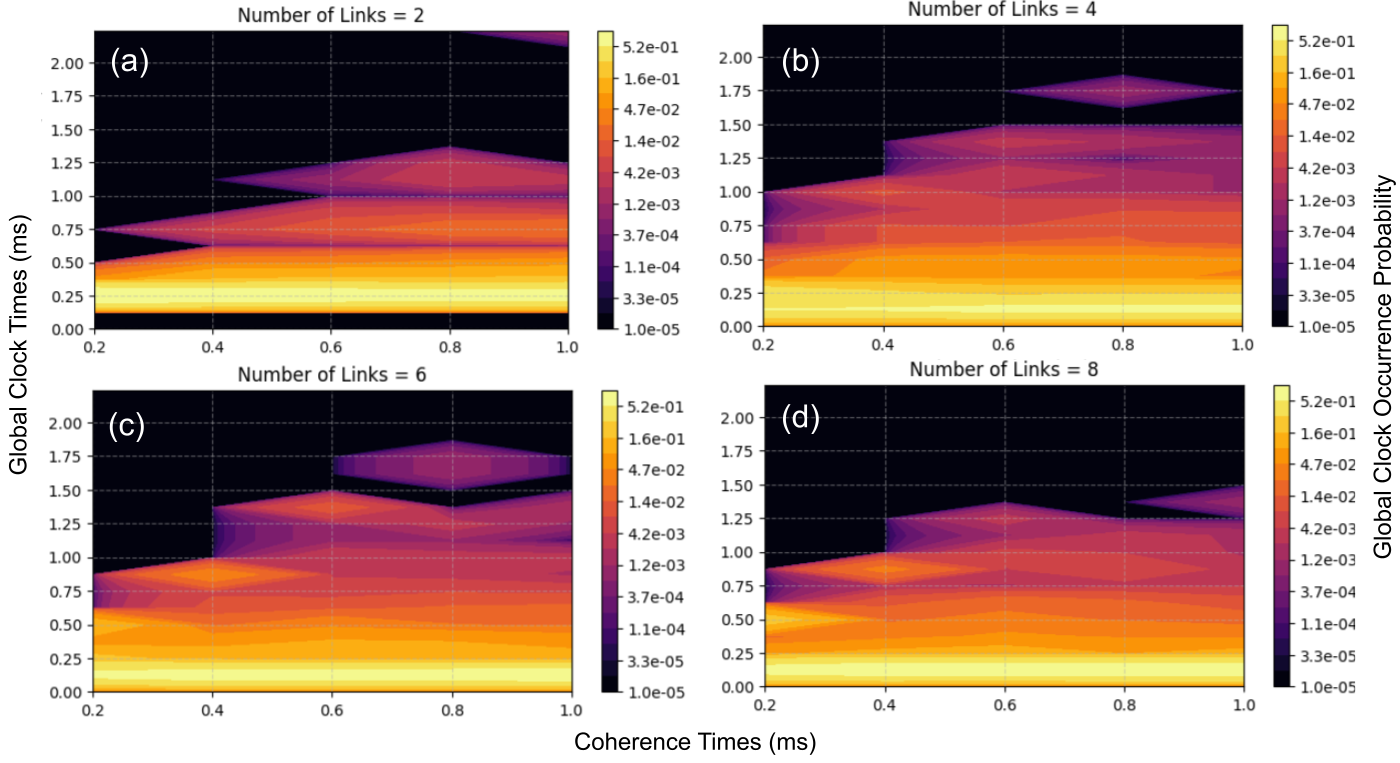}
  \caption{Similar quantity as Fig.~\ref{fig:GCtimevslinks} but with fixed $n$ and varying $\tau_{\scriptscriptstyle mem}$. Contours show distribution of global clock times for 1000 successful entanglement events with fixed number of links: (a) $n=2$, (b) $n=4$, (c) $n=6$, (d) $n=8$. Axes as in Fig.~\ref{fig:GCtimevslinks}. Horizontal axis: coherence time; vertical axis: $T_{\mathrm{GC}}$; Contours are the success probability (log scale). At large $n$, successful trials are rare and conditioned events cluster at shorter $T_{\mathrm{GC}}$ because long waits are unlikely to survive all links.}
  \label{fig:GCtimevscohtime}
  
\end{figure*}

Figs.~\ref{fig:GCtimevslinks} and~\ref{fig:GCtimevscohtime} primarily show the effect that asynchronous completion times $T_{\mathrm{GC}}$ broaden when QMs live longer (i.e. more waiting is tolerable) and when more links must succeed (more sequential steps). This also highlights that very lossy chains or chains with many links are unlikely to succeed.

Our global and local clock system's stochastic nature allows us to analyze the range of global clock times stamped by a successful entanglement distribution event. Figure~\ref{fig:GCtimevslinks} summarizes, for each link count, a histogram of $T_{\mathrm{GC}}$ over 1000 successful trials (not all trials succeed as they are conditioned on success). Coherence times $\tau_{\scriptscriptstyle mem}$ take values of $\in\{0.1, 0.25, 0.5, 1.0\}$~ms are illustrated with $\eta_{\scriptscriptstyle QM}=0.5$. On a logarithmic horizontal axis, longer $\tau_{\scriptscriptstyle mem}$ visibly widens the distribution because the QMs survive the multiple waiting steps. The presence of fewer links modestly narrows the spread because fewer independent stages must align.

For the lowest coherence time of 0.1~ms, variations in global clock time were nonexistent after a certain transit time, as a QM would not be expected to still be in possession of an entangled photon. In contrast, the longest coherence time translated to a larger  variation in the global clock timestamp with times ranging from 0.01 - 2~ms. 

The analysis was also performed for increasing QM coherence time at a fixed number of links. Figure~\ref{fig:GCtimevscohtime} shows the same trend: longer $\tau_{\scriptscriptstyle mem}$ widens the $T_{\mathrm{GC}}$ distribution when $n$ is moderate. For the largest $n$, the histograms appear narrower at long coherence times because conditioning on success selects trials where memories did not fail early, which is an effect due to trial selection.

These figures highlight the functionality and flexibility of our simulation tool. The time at which each QM pair became entangled is provided to be used in a thorough analysis of entanglement distribution rates. The network parameters can be easily modified and can be visualized as a result of the local and global clock system making it a valuable resource for research and optimization.

 As a first treatment, the present study quantifies rates for a homogeneous line segment. The settings assumed in Eqs.~(\ref{eq:multilinkdet})-(\ref{eq:asynchrate}) quantify rates for a homogenous line segment. However, the Monte Carlo treatment presented here could be expanded to a general mesh topology by the addition of routing, cutoffs and multiple paths~\cite{Inesta2023}. This would require introduction of a mechanism to determine which edges of the network are active, and it will be addressed in future studies.

\section*{{Model Optimization and Discussion}}

Future iterations of our in-house simulation tool could include other real-world considerations such as optical mode coupling, filtering, detector efficiency, frequency conversion, and memory write/read efficiencies \cite{Guo2019}.

Furthermore, we would incorporate additional noise sources that increase the accidental coincidence rate and lead to a degradation in Bell state measurement statistics~\cite{Schauer2013}. This would require a full analysis to explicitly propagate imperfections through the swapping hierarchy at each repeater node. Specifically, we would implement factors like non-ideal waveplate settings, beam splitter imbalance, mode mismatch, and timing jitter \cite{ZhongLo2015} that result in a reduction in two-photon interference visibility and therefore reduce the entanglement fidelity passed to deeper nesting levels.

Our model is not limited to fiber systems and can be modified for free-space segments \cite{Krzic2023}. This would involve different parameter modeling to include the effects of atmospheric turbulence, random transverse vector wander, beam spread, and scattering. All these factors can lower link transmissivity and spatial mode overlap at the BSMs, which would reduce the correlations in the final data.

This study assumed ideal single-photon sources, but currently available sources have deficiencies that should be considered in network models. Future modeling could incorporate more practical sources, such as spontaneous parametric down-conversion for entangled pairs \cite{Zang2022}, and account for the impact of multi-pair events. Such multi-photon events could lead to a penalty in the BSM visibility \cite{Fang2025} that would worsen with each additional repeater nesting depth.

\section*{{Conclusion}}

We have presented a stochastic model based on global and local clocks that can be used to predict the entanglement distribution rate in a chain-like network both synchronously and asynchronously. Our model implemented Monte Carlo methods to sample the parameters of QM efficiency and storage lifetime.  When simulating the independent generation of entanglement between adjacent nodes, our model yielded higher distribution rates compared to the synchronous case. Our results provide insight into the performance of a quantum network as a function of the number of repeaters, end-to-end distance~\cite{Inesta2023} and the storage lifetime of the quantum memories~\cite{9495278}. Specifically, the local/global clock system of our model emphasizes analysis of the rate curves in a compact and QM rate-focused layer which can be incorporated into full-stack discrete-event simulators~\cite{Satoh2022, Wu2021}.

Our model's focus on simplicity and real-world visualization makes it a powerful tool for exploring the complex trade-offs inherent in designing practical quantum communication networks. The approach implements global and local simulation clocks to track quantum memory lifetimes, a key feature that allows for easy visualization of successful asynchronous entanglement distribution events. This provides a clear and intuitive way to understand network behavior when QMs are used, making our simulator a useful complement to both research and education. Further, the in-house simulation algorithm is a valuable addition to the existing family of quantum network simulators \cite{DiAdamo2021_2, Satoh2022, Wu2021}.

\section*{{Acknowledgments}}

The authors would like to thank Kenny Gregory for fruitful discussions. We acknowledge the NSERC USRA and NSERC Discovery program for funding support.

\bibliographystyle{unsrt}
\bibliography{mainbib}

\end{document}